\documentclass[aps,pra,showpacs,amsmath,amssymb,amsfonts,superscriptaddress,twocolumn,lengthcheck,longbibliography]{revtex4-2}
\usepackage{amsmath}
\usepackage{graphicx}
\usepackage{subfigure}
\usepackage{verbatim}
\usepackage{dcolumn}
\usepackage{bm}
\usepackage{epsf}
\usepackage{xcolor}
\usepackage{hhline}
\usepackage{txfonts}
\usepackage{float}
\usepackage{enumerate}
\usepackage{bbm}
\usepackage{mathrsfs}
\usepackage{lipsum}
\usepackage{appendix}
\usepackage{braket}
\usepackage{float}
\usepackage{physics}
\usepackage[colorlinks=true,urlcolor=blue,citecolor=blue]{hyperref}

\begin{document}

	\title{ Preparing two-mode magnonic Schrödinger cat states in a cavity-magnon-qubit system} 
	\author{Gen Li}
	\thanks{These authors contributed equally to this work.}
	\affiliation{Zhejiang Key Laboratory of Micro-Nano Quantum Chips and Quantum Control, 	and School of Physics, Zhejiang University, Hangzhou 310027, China}

	\author{Gang Liu}
	\thanks{These authors contributed equally to this work}\thanks{gangliuphys@gmail.com}
	\affiliation{Zhejiang Key Laboratory of Micro-Nano Quantum Chips and Quantum Control, 	and School of Physics, Zhejiang University, Hangzhou 310027, China}

	\author{Rong-Can Yang}
	\affiliation{Zhejiang Key Laboratory of Micro-Nano Quantum Chips and Quantum Control, 	and School of Physics, Zhejiang University, Hangzhou 310027, China}

	\author{Jie Li}\thanks{jieli007@zju.edu.cn}
	\affiliation{Zhejiang Key Laboratory of Micro-Nano Quantum Chips and Quantum Control, 	and School of Physics, Zhejiang University, Hangzhou 310027, China}

\begin{abstract}
The cavity-magnon-qubit system has recently been demonstrated as a new platform for preparing macroscopic quantum states in magnonic systems. Here,  we propose to prepare a two-mode magnonic cat state, which is also a non-Gaussian entangled state, based on this practical system involving two yttrium-iron-garnet (YIG) spheres and a superconducting qubit coupled to a common  microwave cavity.  By adiabatically eliminating the cavity and resonantly driving the qubit, an effective magnon-qubit conditional-displacement interaction is achieved.  Further working in the magnon-magnon strong-coupling regime and considering two identical magnon frequencies and coupling strengths to the cavity, two hybridized magnon modes are formed, of which the bright mode is prepared in a cat state after a projective measurement on the qubit, while the dark mode remains in its initial vacuum state. Such a state corresponds to a two-mode cat state of two original magnon modes, which share strong non-Gaussian entanglement.  We also discuss practical dissipation and dephasing effects on the cat state. The results indicate that strong nonclassicality and non-Gaussian entanglement are present in the two-mode cat state using fully feasible parameters.
\end{abstract}
\maketitle

\section{INTRODUCTION}

Schrödinger cat states, which correspond to a superposition of two coherent states with opposite phases, have attracted considerable interest as a typical representative of macroscopic quantum states and play an important role in testing the limit of quantum theory~\cite{RevModPhys.75.715, NaturePhysics2014}. They have been extensively studied in both theory and experiment. To date, different mechanisms have been proposed to generate cat states exploiting, e.g., conditional displacement~\cite{PhysRevLett.90.027903, PhysRevLett.116.163602, PhysRevA.93.033853}, excitation subtraction~\cite{PhysRevA.101.033812, PhysRevA.101.063834, li2026}, driven-dissipative mechanism~\cite{SR2016, PhysRevA.110.013711, zhgm-p3ss}, dispersive coupling~\cite{Savage:OL1990, PhysRevA.45.5193, PhysRevLett.77.4887}, etc.
They have been experimentally realized in trapped ions~\cite{science.272.5265.1131, Nature2005}, microwave photons~\cite{science.1243289}, optical photons~\cite{science.1122858, PhysRevLett.97.083604, PhysRevLett.115.023602}, as well as a bulk mechanical oscillator~\cite{science.adf7553}.
Beyond a single-mode cat state, extending cat states to multi-mode systems enables enhanced quantum sensing and a larger quantum-information capacity. In particular, two-mode cat states, namely coherent-state superpositions across two bosonic modes~\cite{PhysRevA.45.6811, Gerry1997, IOP2012, science.aaf2941}, are an important class of non-Gaussian entangled states and a valuable resource for various quantum information protocols -- they can improve the fidelity in quantum communication in the presence of loss channels~\cite{NaturePhysics2009, PhysRevLett.105.160501}, and enhance quantum metrology and sensing, e.g., phase estimation~\cite{PhysRevLett.107.083601} and dark matter detection~\cite{wbhn-v1sw}. 

In recent years, hybrid systems based on magnons in ferromagnetic crystals, such as yttrium-iron-garnet (YIG) spheres, have become a new arena to prepare macroscopic quantum states, study rich nonlinear effects, and develop novel quantum technologies~\cite{Lachance-Quirion_2019, YUAN20221, ZARERAMESHTI20221, Zuo_2024}. Owing to their great tunability and excellent ability to couple with diverse physical systems, including microwave photons~\cite{PhysRevLett.111.127003, PhysRevLett.113.083603, PhysRevLett.113.156401}, optical photons~\cite{PhysRevB.93.174427, PhysRevLett.116.223601,PhysRevLett.117.123605, PhysRevLett.117.133602}, phonons~\cite{PhysRevLett.121.203601, sciadv.1501286, PhysRevX.11.031053, PhysRevLett.129.123601, PhysRevLett.129.243601}, and superconducting qubits~\cite{science.aaa3693, sciadv.1603150, science.aaz9236, PhysRevLett.130.193603,NatureCommunications2026}, magnonic systems hold great potential in quantum information science and technology and quantum sensing: they are an ideal platform for building hybrid quantum systems~\cite{ShenNatureCommunications2025} and quantum networks~\cite{PRXQuantum.2.040344}, realizing fault-tolerant quantum computation~\cite{lu2026}, and searching for dark-matter axions~\cite{PhysRevLett.124.171801}. In particular, the magnon-qubit system, established by their coupling to a common microwave cavity, shows an exceptional ability to prepare magnonic quantum states~\cite{PhysRevLett.130.193603,NatureCommunications2026}, benefiting from the strong anharmonicity of the superconducting qubit, which brings in necessary nonlinearity for creating quantum states.   Theoretical proposals have indicated the potential of the magnon-qubit system in preparing different kinds of magnonic quantum states, including squeezed states~\cite{PhysRevA.108.063703, d2st-rr91, NatureCommunications2026}, Schrödinger cat states~\cite{PhysRevA.110.013711, zhgm-p3ss, PhysRevLett.129.037205, PhysRevA.107.023709, PhysRevA.107.033702, PhysRevA.110.053710, sw7f-syvg}, GHZ states~\cite{PhysRevA.105.022624}, and NOON states~\cite{PhysRevA.107.013702}. We note that previous studies~\cite{PhysRevA.110.013711, zhgm-p3ss, PhysRevLett.129.037205, PhysRevA.107.023709, PhysRevA.107.033702, PhysRevA.110.053710, sw7f-syvg} focused on the generation of single-mode magnonic cat states. The protocol for generating multi-mode cat states, however, is still lacking. 

Here, we propose to generate a two-mode magnonic cat state in a cavity-magnon-qubit system.  The superconducting qubit is simultaneously coupled to two magnon modes of two YIG spheres via the mediation of a microwave cavity.  By applying a resonant microwave drive to the qubit, we realize a conditional-displacement interaction between the qubit and each magnon mode. Further assuming two identical magnon frequencies and coupling strengths to the qubit and working in the strong-coupling regime, a projective measurement on the qubit leads one hybridized magnon mode to a cat state, while the other hybridized dark mode remains in its initial state. This corresponds to a two-mode cat state of two original magnon modes in two  YIG spheres.  We further investigate the effects of various dissipations and dephasing on our protocol and characterize the non-Gaussian entanglement shared between the two YIG spheres.  Finally, we discuss how to verify the  cat state by measuring a joint two-mode Wigner function exploiting the magnon-qubit dispersive coupling.

The paper is organized as follows. In Sec.~\ref{S2}, we introduce the system and derive an effective conditional-displacement interaction between the qubit and the magnon modes. In Sec~\ref{S3},  we explicitly show how to generate a two-mode magnonic cat state by using the conditional-displacement interaction and a qubit projective measurement. In Sec.~\ref{S4}, we analyze the dissipation and dephasing effects on the cat state and characterize the non-Gaussian entanglement shared between the two magnon modes. Finally, we draw the conclusions in Sec.~\ref{S5}.

\begin{figure}[t]
	\centering 
	\hskip-0.7cm\includegraphics[angle=0,width=0.9\linewidth]{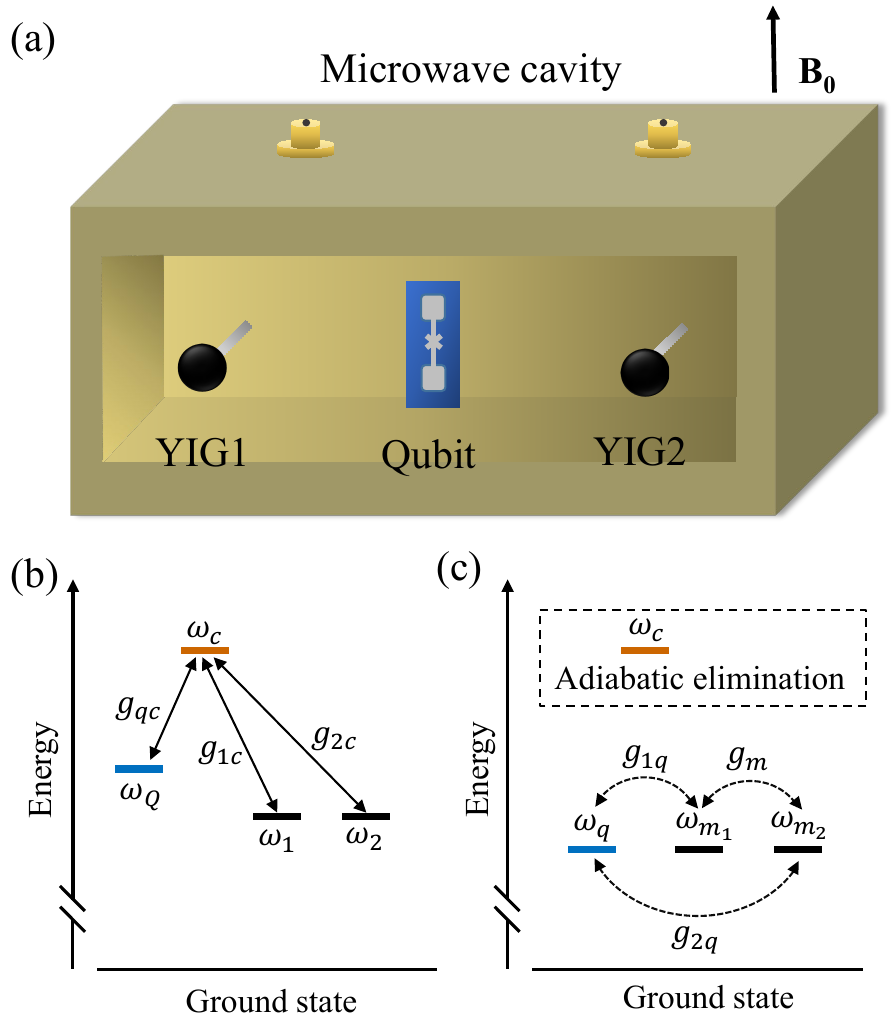}
	\renewcommand{\figurename}{FIG}
	\caption{(a) The cavity-magnon-qubit system, involving a microwave cavity, two YIG spheres, and a superconducting qubit.  A bias magnetic field $B_0$ is applied to have two identical magnon frequencies.  {(b) The two magnon modes (with frequencies $\omega_{1,2}$) and the qubit (with frequency $\omega_Q$) are nearly resonant and directly coupled to a far-detuned cavity mode (with frequency $\omega_c$). } There are no direct couplings among the magnon-magnon-qubit system.  (c) Adiabatic elimination of the microwave cavity leads to effective frequencies of the magnon modes ($\omega_{m_{1}}$ and $\omega_{m_{1}}$) and the qubit ($\omega_{q}$), and effective couplings among the two magnon modes and qubit.  A resonant case is adopted in our protocol, where $\omega_{m_{1}}=\omega_{m_{2}}=\omega_{q}$ and the qubit is resonantly driven by a microwave field.}
	\label{Fig1}
\end{figure}

 \section{The System and Effective Hamiltonian}\label{S2}

The proposed system consists of two YIG spheres, a superconducting qubit, and a three-dimensional microwave cavity, as depicted in Fig.~\ref{Fig1}(a). The YIG spheres are positioned inside a bias magnetic field $ B_{0} $, each supporting a magnon mode, e.g., the Kittel mode. The magnon modes couple to the cavity mode via the magnetic dipole interaction, and the superconducting qubit couples to the cavity mode via the electric dipole interaction. There are no direct couplings among the two magnon modes and the qubit. The Hamiltonian of the whole system reads (setting $ \hbar = 1$)
\begin{equation}\label{initial}
\begin{split}
  H_1 = &\;\;  \omega_c c^{\dagger} c + \frac{\omega_{Q}}{2} \sigma_z + \omega_1  m_{1}^{\dagger} m_{1} + \omega_2 m_{2}^{\dagger} m_{2}\\
   & + g_{1c}\left(m_{1}^{\dagger} c + m_{1} c^{\dagger} \right) + g_{2c} \left( m_{2}^{\dagger} c + m_{2} c^{\dagger} \right) \\
   & + g_{q c} \left(\sigma_{+}c + \sigma_{-}c^{\dagger} \right),
\end{split}
\end{equation} 
where $ c $ ($ c^{\dagger}  $) and $ m_{i} $ ($ m^{\dagger}_{i}  $)  are the annihilation (creation) operators of the cavity mode and the $ i $th magnon mode ($ i $ = 1,2) with frequencies $ \omega_{c} $ and $ \omega_{i} $, respectively. $ \sigma_{j} $ ($ j = x, y, z $) denote the Pauli operators and $ \sigma_{\pm} = (\sigma_{x} \pm i \sigma_{y})/2 $ are the ladder operators of the qubit with resonant frequency $ \omega_{Q} $. $ g_{ic} $ ($ g_{qc} $) represents the coupling strength between the $i$th magnon mode (the qubit) and the cavity mode, as illustrated in Fig.~\ref{Fig1}(b).

When the cavity is far detuned from both the qubit and the magnon modes, i.e., $ \Delta_{Qc} = \omega_{c} - \omega_{Q} \gg g_{qc} $ and $ \Delta_{ic} = \omega_{c} - \omega_{i} \gg g_{ic} $,  the magnon modes and the qubit are coupled via the exchange of virtual photons of the microwave cavity. This allows us to adiabatically eliminate the cavity mode and obtain effective couplings between the magnon modes and the qubit, as illustrated in Fig.~\ref{Fig1}(c). Consequently, we obtain the following effective Hamiltonian of the magnon-qubit system (Appendix \ref{appa}): 
\begin{equation}\label{hh}
\begin{split}
  H_2 =& \;\; \frac{\omega_q}{2} \sigma_z + \omega_{m_{1}}  m_{1}^{\dagger} m_{1} + \omega_{m_{2}} m_{2}^{\dagger} m_{2}\\
  &+ g_{1q} \left(\sigma_{+}m_{1} + \sigma_{-} m_{1}^{\dagger}  \right) + g_{2q} \left( \sigma_{+} m_{2} + \sigma_{-} m_{2}^\dagger  \right)\\
  &+ g_{m} \left(m_{1}^{\dagger} m_{2} + m_{1} m_{2}^{\dagger} \right),
\end{split}
\end{equation}
where $ \omega_{q} = \omega_{Q} - g_{qc}^2/\Delta_{Qc}  $ and $ \omega_{m_{i}} = \omega_{i} - g_{ic}^2/\Delta_{ic} $ correspond to the effective frequencies of the qubit and the $ i $th magnon mode, respectively. {The effective magnon-qubit coupling is  $ g_{iq} = -g_{ic} g_{qc}(\Delta_{ic} + \Delta_{Qc})/(2 \Delta_{ic} \Delta_{Qc}) $, and the magnon-magnon coupling is $ g_{m} = -g_{1c} g_{2c}(\Delta_{1c} + \Delta_{2c})/(2 \Delta_{1c} \Delta_{2c})$}. Such cavity-mediated couplings  enable controllable and strong long-range interactions between the magnon and qubit systems~\cite{science.aaa3693, sciadv.1603150, science.aaz9236, PhysRevLett.130.193603,NatureCommunications2026} with otherwise weak direct coupling. 

The qubit is driven by a coherent microwave field with frequency $ \omega_{d} $ and driving strength $ \Omega$. In the frame rotating at the drive frequency, 
the Hamiltonian is given by
\begin{equation}
\begin{split}
	H_3 
	&= \frac{\Delta_{q}}{2} \sigma_{z} + \Delta_{m_{1}} m_{1}^{\dagger} m_{1} + \Delta_{m_{2}} m_{2}^{\dagger} m_{2}\\
	&+ g_{1q} \left(\sigma_{+}m_{1} + \sigma_{-} m_{1}^{\dagger} \right) + g_{2q} \left( \sigma_{+} m_{2} + \sigma_{-} m_{2}^{\dagger}  \right) \\
  	&+ g_{m} \left( m_{1}^{\dagger} m_{2} + m_{1} m_{2}^{\dagger} \right) + \Omega \sigma_{x}, 
\end{split}
\end{equation}
where $ \Delta_{q} = \omega_{q} - \omega_{d} $, $ \Delta_{m_{1}} = \omega_{m_{1}} - \omega_{d} $, and $ \Delta_{m_{2}} = \omega_{m_{2}} - \omega_{d} $. For the resonant case $ \Delta_{q} =  \Delta_{m_{1}}  =  \Delta_{m_{2}}  =0$, the Hamiltonian simplifies to
\begin{equation}
\begin{split}
  H_4 &=  g_{1q} \left(\sigma_{+}m_{1} + \sigma_{-} m_{1}^{\dagger} \right) + g_{2q} \left( \sigma_{+} m_{2} + \sigma_{-} m_{2}^{\dagger} \right) \\
  & + g_{m} \left(m_{1}^{\dagger} m_{2} + m_{1} m_{2}^{\dagger} \right) +  \Omega \sigma_{x}.
\end{split}
\end{equation}
The above Hamiltonian in the interaction picture with respect to $ H_{\text{0}} = \Omega \sigma_{x} $ becomes 
\begin{equation}\label{RWA}
\begin{split}
  H_5 =&\;\; \frac{g_{1q}}{2} \left( m_{1} + m_{1}^{\dagger} \right) \sigma_{x} + \frac{g_{2q}}{2} \left( m_{2} + m_{2}^{\dagger} \right) \sigma_{x}\\
  &+ i \frac{g_{1q}}{2} \left( m_{1} - m_{1}^{\dagger} \right) \left[  \sigma_{y} \cos(2 \Omega t) + \sigma_{z} \sin(2 \Omega t)  \right] \\
  & + i \frac{g_{2q}}{2} \left( m_{2} - m_{2}^{\dagger} \right) \left[  \sigma_{y} \cos(2 \Omega t) + \sigma_{z} \sin(2 \Omega t)  \right] \\
  & + g_{m} \left(m_{1}^{\dagger} m_{2} + m_{1} m_{2}^{\dagger} \right).
\end{split}
\end{equation}
For a sufficiently strong drive field with $\Omega \gg g_{1q}, g_{2q} $, which allows us to make the rotating-wave approximation (RWA) by neglecting the fast-oscillating terms in Eq.~\eqref{RWA}, we obtain
\begin{equation}\label{hheff}
\begin{split}
  H_6 =&\;\; \frac{g_{1q}}{2} \left( m_{1} + m_{1}^{\dagger} \right) \sigma_{x}  + \frac{g_{2q}}{2}  \left( m_{2} + m_{2}^{\dagger} \right) \sigma_{x}\\
  & + g_{m} \left(m_{1}^{\dagger} m_{2} + m_{1} m_{2}^{\dagger} \right).
\end{split}
\end{equation}
{To achieve strong non-Gaussian entanglement between the two magnon modes, we work in the strong-coupling regime, where the two magnon modes are hybridized,} and the two hybridized modes (upper- and lower-branch) are denoted by the annihilation operators $ U = ( m_{1} + m_{2} )/\sqrt{2} $ and $ L = ( m_{1} - m_{2} )/\sqrt{2} $, respectively. The Hamiltonian~\eqref{hheff} then can be rewritten as 
\begin{equation}\label{777}
\begin{split}
   H_7 &= g_{m}U^{\dagger} U - g_{m}L^{\dagger} L\\
   & + \frac{g_{1q} + g_{2q}}{2\sqrt{2}} \left( U + U^{\dagger} \right) \sigma_{x}  + \frac{g_{1q} - g_{2q}}{2\sqrt{2}} \left( L + L^{\dagger} \right) \sigma_{x}.
\end{split}
\end{equation}
We further consider the symmetric situation, i.e., the two magnon modes have identical frequencies and coupling strengths with the cavity,  which lead to identical effective couplings $ g_{1q} = g_{2q} \equiv g $. This results in the decoupling between the qubit and the lower hybridized mode (LHM) due to the complete destructive interference between the two coupling channels (this mode is thus called the ``dark mode''). 
Consequently, the Hamiltonian~\eqref{777} reduces to
\begin{equation}\label{heff}
\begin{split}
  H_8=  g_{m}U^{\dagger} U - g_{m}L^{\dagger} L  + \frac{ g }{\sqrt{2}} \left( U + U^{\dagger} \right) \sigma_{x}.
\end{split}
\end{equation}
Since the LHM is decoupled from the system, if initialized in the vacuum state, it remains in the vacuum state during the evolution. By contrast, the upper hybridized mode (UHM) couples to the qubit through a conditional-displacement interaction, which, we will show in Sec.~\ref{S3}, is a key ingredient in our protocol for preparing two-mode magnonic cat states.

\section{Generating TWO-MODE CAT STATES}\label{S3}

In what follows, we explicitly show how the effective Hamiltonian~\eqref{heff} can be used to generate two-mode cat states in our system.
The Hamiltonian~\eqref{heff}, excluding the dark mode, corresponds to the following unitary evolution operator ${\cal U}(t)$  (Appendix \ref{appb}):
\begin{equation}\label{Ut}
\begin{split}
	{\cal U}(t) = \exp \left[ i \Phi(t) \right]  \exp \left\{ \left[  \alpha(t) U^{\dagger} - \alpha^{*}(t) U \right] \sigma_{x} \right\},
\end{split}
\end{equation}
where $ \Phi(t) = g^2 (g_{m} t - \sin g_{m} t)/(2 g_{m}^2) $ represents a global phase, {and $ \alpha(t) = g[1 - \exp(i g_{m} t)]/(\!\sqrt{2} g_{m}) $ denotes the amount of displacement onto the UHM. }
Equation~\eqref{Ut} indicates a displacement of the hybridized mode conditioned on the qubit state: the two eigenstates of $\sigma_{x}$ with eigenvalues $\pm1$ correspond to the displacement of the hybridized mode in opposite directions.
We assume that the qubit is initialized in the ground state, $\ket{ g }  = (\ket{ + } - \ket{ - })/\!\sqrt{2}$, i.e., a superposition of two eigenstates of $\sigma_{x}$, and the magnon modes are initialized in their vacuum state $\ket{0}$. 
The system evolves into the following state at time $t$ governed by the evolution operator ${\cal U}(t)$:
\begin{equation}
\begin{split}
	\ket{ \psi(t) } = \frac{ e^{i \Phi(t)}  }{\sqrt{2}} \left( \ket{ \alpha(t) }_{U} \otimes \ket{+} - \ket{ -\alpha(t) }_{U} \otimes \ket{-} \right) \otimes \ket{ 0 }_{L},
\end{split}
\end{equation}
where $\ket{ \alpha(t) }_{U}$ ($\ket{ 0 }_{L}$) denotes a coherent (vacuum) state of the UHM (LHM).
Projecting the qubit onto the ground state $\ket{g}$ leads the UHM to an even cat state, while the LHM remains in the vacuum state, i.e.,
\begin{equation}\label{cat_0}
\begin{split}
	\ket{ \psi(t) }_{g} = \frac{1}{ \mathcal{N}_e } \left( \ket{\alpha(t)}_{U} + \ket{ - \alpha(t) }_{U} \right) \otimes \ket{ 0 }_{L},
\end{split}
\end{equation}
with $ \mathcal{N}_e = \sqrt{2 + 2\exp( - 2 \abs{\alpha(t)}^2 )}  $ being the normalization factor. If, instead, the qubit is projected onto the excited state $\ket{e}$, the UHM is prepared in an odd cat state, i.e.,
\begin{equation}\label{cat_1}
\begin{split}
	\ket{ \psi(t) }_{e} = \frac{1}{ \mathcal{N}_o } \left( \ket{\alpha(t)}_{U} - \ket{ - \alpha(t) }_{U} \right) \otimes \ket{ 0 }_{L},
\end{split}
\end{equation}
with $ \mathcal{N}_o = \sqrt{2 - 2\exp( - 2 \abs{\alpha(t)}^2 )}  $.
The cat states~\eqref{cat_0} and~\eqref{cat_1}  are expressed in terms of two hybridized magnon modes, 
and they correspond to the following two-mode cat states in terms of two original magnon modes (unnormalized):
\begin{equation}\label{target}
\begin{split}
	\ket{ \psi(t) }_{\pm} =  \left( \ket{\alpha'(t), \alpha'(t)}_{m_1,m_2} \pm \ket{ - \alpha'(t), - \alpha'(t) }_{m_1,m_2} \right),
\end{split}
\end{equation}
where $  \alpha'(t) = \alpha(t)/\sqrt{2} $. 
The above states represent a coherent superposition of two macroscopically distinguishable states involving two magnon modes,  
which are also non-Gaussian entangled states. Without loss of generality, hereafter we consider the case where the qubit is projected onto the ground state.

\begin{figure}[t]
	\centering
	\includegraphics[angle=0,width=0.95\linewidth]{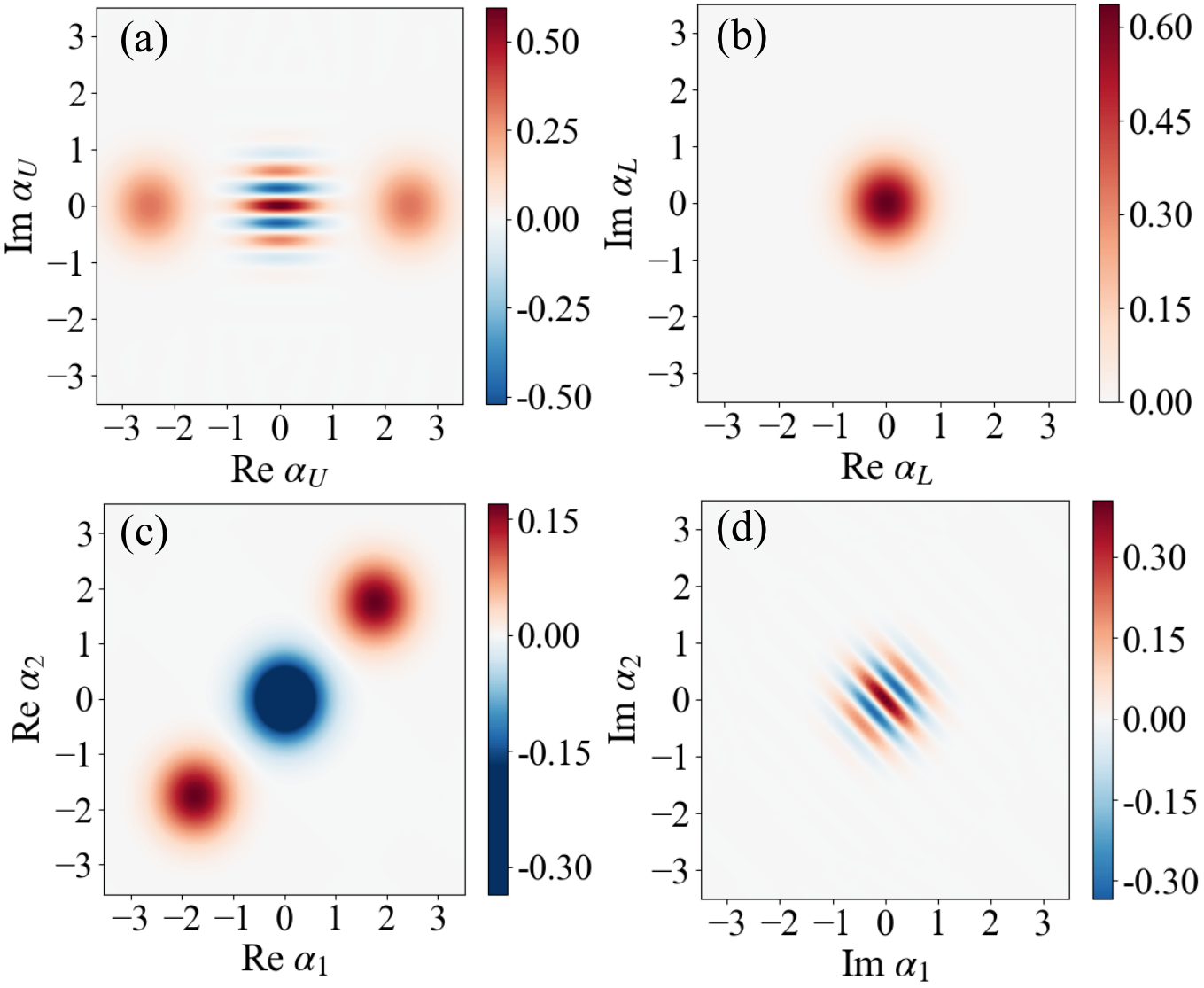}
	\renewcommand{\figurename}{FIG}
	\caption{Wigner function of (a) the UHM and (b) the LHM. Two-dimensional cuts of the two-mode Wigner function versus (c) Re $\alpha_{1}$ and Re $\alpha_{2}$  (${\rm Im} \, \alpha_{1}= {\rm Im} \, \alpha_{2}= 0.3$), and (d) Im $\alpha_{1}$ and Im $\alpha_{2}$ (${\rm Re} \, \alpha_{1}= {\rm Re} \, \alpha_{2}= 0$). See text for the parameters.} 
	\label{Fig3}
\end{figure}

 In Figs.~\ref{Fig3}(a)-(b), we plot the Wigner function of the two hybridized magnon modes (UHM and LHM) at time $t = \pi/g_{m}$, under the evolution governed by the Hamiltonian~\eqref{heff}.  At this time, the maximum displacement $ |\alpha(t) | =  \sqrt{2} g /g_{m} \approx 2.48 $ is achieved.  Clearly, the Wigner function of the UHM exhibits pronounced interference fringes between two coherent components, at which the Wigner displays negative-value distributions -- a typical characteristic of the cat state. In contrast, the Wigner function of the LHM corresponds to the vacuum state because it is a dark mode and initially in the vacuum state. These results are consistent with Eq.~\eqref{cat_0}, thereby confirming the validity of our analytical results. 
We have used fully feasible parameters~\cite{science.aaz9236, PhysRevLett.130.193603,NatureCommunications2026}: $ \omega_{c}/2\pi = 6.328$~GHz, $ \omega_{1,2}/2\pi = 5.914$~GHz, $\omega_{Q} = 5.927$~GHz, $ g_{1c}/2\pi = g_{2c}/2\pi = 51$~MHz, $ g_{qc}/2\pi = 88$~MHz, and $\Omega/2\pi = 100$~MHz, which yield the following effective frequencies and couplings $ \omega_{m_{1,2}}/2\pi = \omega_{q}/2\pi = 5.908$~GHz, {$ |g_{1q}|/2\pi = |g_{2q}|/2\pi = 11.004$~MHz, and $ |g_{m}|/2\pi = 6.281$~MHz.}

As explained above, the cat state of the UHM corresponds to a two-mode cat state of the original magnon modes, which can be characterized by a joint Wigner function, defined as
\begin{equation}
\begin{split}
	W(\alpha_{1},\alpha_{2}) = \frac{4}{\pi^2} \Tr \left[ \rho_{12}  D_{1}(\alpha_{1}) D_{2}(\alpha_{2}) P D_{1}(-\alpha_{1}) D_{2}( - \alpha_{2}) \right],   
\end{split}
\end{equation}
where $ \rho_{12}  $ is the reduced density matrix of the two magnon modes, $ P \equiv P_{\text{1}} P_{\text{2}} = e^{i \pi m_{1}^{\dagger} m_{1}} e^{i \pi m_{2}^{\dagger} m_{2}} $ is  the joint parity operator, and $ D_{i}(\alpha_{i}) $ denotes the displacement operator associated with the $i$th magnon mode. Since the two-mode Wigner function $W(\alpha_{1},\alpha_{2})$ is four-dimensional, in Figs.~\ref{Fig3}(c)-(d) we present the two-dimensional cuts of the Wigner function at time $t = \pi/g_{m}$. 
In the Re $\alpha_{1}$-Re $\alpha_{2}$ panel (Fig.~\ref{Fig3}(c)), the Wigner function exhibits two positive Gaussian wave packets and a negative wave packet centered at the origin, while the interference fringes around the origin with pronounced negativity are shown in the Im $\alpha_{1}$-Im $\alpha_{2}$ panel (Fig.~\ref{Fig3}(d)). The two sections of the joint Wigner function reveal not only the nonclassicality but also the quantum correlation of the two-mode cat state, which, as will be shown in Sec.~\ref{S4}, shares strong non-Gaussian entanglement.

\begin{figure}[t]
	\centering
	\includegraphics[angle=0,width=0.95\linewidth]{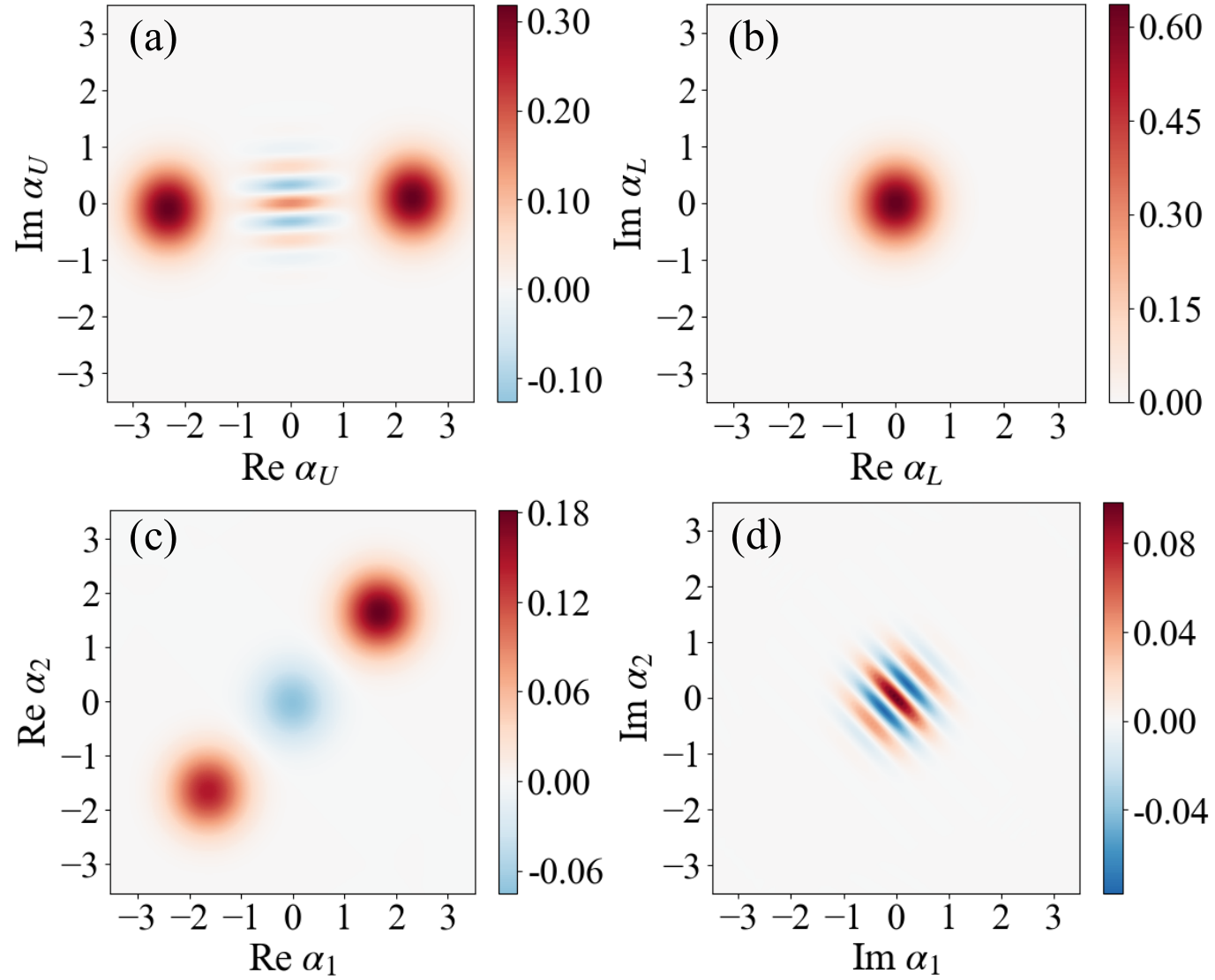}
	\renewcommand{\figurename}{FIG}
	\caption{Wigner function of (a) the UHM and (b) the LHM. Two-dimensional cuts of the two-mode Wigner function versus (c) Re $\alpha_{1}$ and Re $\alpha_{2}$  (${\rm Im} \, \alpha_{1}= {\rm Im} \, \alpha_{2}= 0.3$), and (d) Im $\alpha_{1}$ and Im $\alpha_{2}$ (${\rm Re} \, \alpha_{1}= {\rm Re} \, \alpha_{2}= 0$).  We take $ \kappa_{1}/2\pi = \kappa_{2}/2\pi = 0.5$ MHz~\cite{ShenNatureCommunications2025}, and $ \gamma_{q}/2\pi = \gamma_{\phi}/2\pi = 5 $ kHz. The other parameters are the same as those in Fig.~\ref{Fig3}.} 
	\label{Fig4}
\end{figure}

 \section{Effects of system dissipations and dephasing}\label{S4}

We have analytically shown that our protocol can generate a dynamical two-mode magnonic cat state without considering any dissipation and dephasing. In practice, the system is inevitably affected by losses and noises. In this section, we adopt the master equation approach to numerically investigate how these effects affect our protocol. We start from the magnon-qubit Hamiltonian~\eqref{hh} after the elimination of the cavity, and the corresponding master equation reads
\begin{equation}\label{meq}
\begin{split}
	\frac{\mathrm{d} \rho}{\mathrm{d} t} &= - i \left[ H_2, \rho \right] + \frac{\gamma_{q} ( \bar{n}_{q} + 1 )}{2} \mathcal{L}[\sigma_{-}] \rho + \frac{\gamma_{q} \bar{n}_{q}}{2} \mathcal{L}[\sigma_{+}] \rho  \\
	&+ \frac{\gamma_{\phi}}{4} \mathcal{L}[\sigma_{z}] \rho
	+ \sum_{i = 1}^{2} \frac{\kappa_{i} ( \bar{n}_{i} {+} 1)}{2} \mathcal{L}[m_{i}] \rho +\frac{\kappa_{i} \bar{n}_{i} }{2} \mathcal{L}[m_{i}^{\dagger} ] \rho, 
\end{split}
\end{equation}
where $ \rho $ is the density operator of the magnon-qubit system, and $\mathcal{L}[o] \rho = 2 o \rho o^{\dagger} - o^{\dagger} o \rho - \rho o^{\dagger} o $ is the standard Lindblad superoperator with $ o = \{\sigma_{-}, \sigma_{z}, m_{1}, m_{2}\} $. Here, $ \kappa_{i} $ ($ \gamma_{q} $) is the dissipation rate of the $ i $th magnon mode (the qubit), $ \gamma_{\phi} $ signifies the pure dephasing rate of the qubit, and $\bar{n}_{i(q)} = \left[\exp(\hbar \omega_{m_i(q)} / k_{B} T) - 1\right]^{-1}$ is the mean thermal excitation number of the $ i $th magnon mode (the qubit). For a low bath temperature, e.g., a few tens of mK, the mean thermal excitations of the qubit and magnon modes with frequency about 5.9~GHz are essentially zero, $\bar{n}_{i(q)} \approx 0$, which leads to a simpler form of the master equation
\begin{equation}\label{meq2}
	\frac{\mathrm{d} \rho}{\mathrm{d} t} = - i \left[ H_2, \rho \right] + \frac{\gamma_{q}}{2} \mathcal{L}[\sigma_{-}] \rho + \frac{\gamma_{\phi}}{4} \mathcal{L}[\sigma_{z}] \rho  +  \sum_{i = 1}^{2} \!\frac{\kappa_{i}}{2} \mathcal{L}[m_{i}] \rho . 
\end{equation}
 In the rotating frame, we derive the following effective master equation corresponding to the Hamiltonian~\eqref{hheff}, given by (Appendix \ref{appc})
\begin{equation}\label{meff}
\begin{split}
	\frac{\mathrm{d} \tilde{\rho}}{\mathrm{d} t} =& - i \left[ H_6, \tilde{\rho} \right] + \frac{\kappa_{1}}{2} \mathcal{L}[m_{1}] \tilde{\rho} + \frac{\kappa_{2}}{2} \mathcal{L}[m_{2}] \tilde{\rho} \\
	&+ \frac{\gamma_{q}}{8} \mathcal{L}[\sigma_{x}] \tilde{\rho} + \frac{\gamma_{q}}{16} \mathcal{L}[\sigma_{y}] \tilde{\rho} + \frac{\gamma_{q}}{16} \mathcal{L}[\sigma_{z}] \tilde{\rho} \\
	&+ \frac{\gamma_{\phi}}{8} \mathcal{L}[\sigma_{y}] \tilde{\rho} + \frac{\gamma_{\phi}}{8} \mathcal{L}[\sigma_{z}] \tilde{\rho},
\end{split}
\end{equation}
where $ \tilde{\rho} $ denotes the density matrix in the rotating frame.

 In Fig.~\ref{Fig4}, we present the corresponding Wigner functions by including the dissipation and dephasing effects. Note that the joint Wigner functions of the original magnon modes (Figs.~\ref{Fig4} (c)-(d)) are obtained by numerically solving the master equation~\eqref{meff}, whereas the Wigner functions of the hybridized modes (Figs.~\ref{Fig4} (a)-(b)) are obtained by solving the accordingly transformed master equation.  
 Clearly, the dissipation and dephasing reduce the negativity of the Wigner functions and the visibility of the interference fringes, i.e., the nonclassicality of the two-mode cat state.
 In contrast, the Wigner function of the dark hybridized mode in Fig.~\ref{Fig4}(b) is unchanged as it remains in the vacuum state.

\begin{figure}[t]
	\centering
	\includegraphics[angle=0,width=0.95\linewidth]{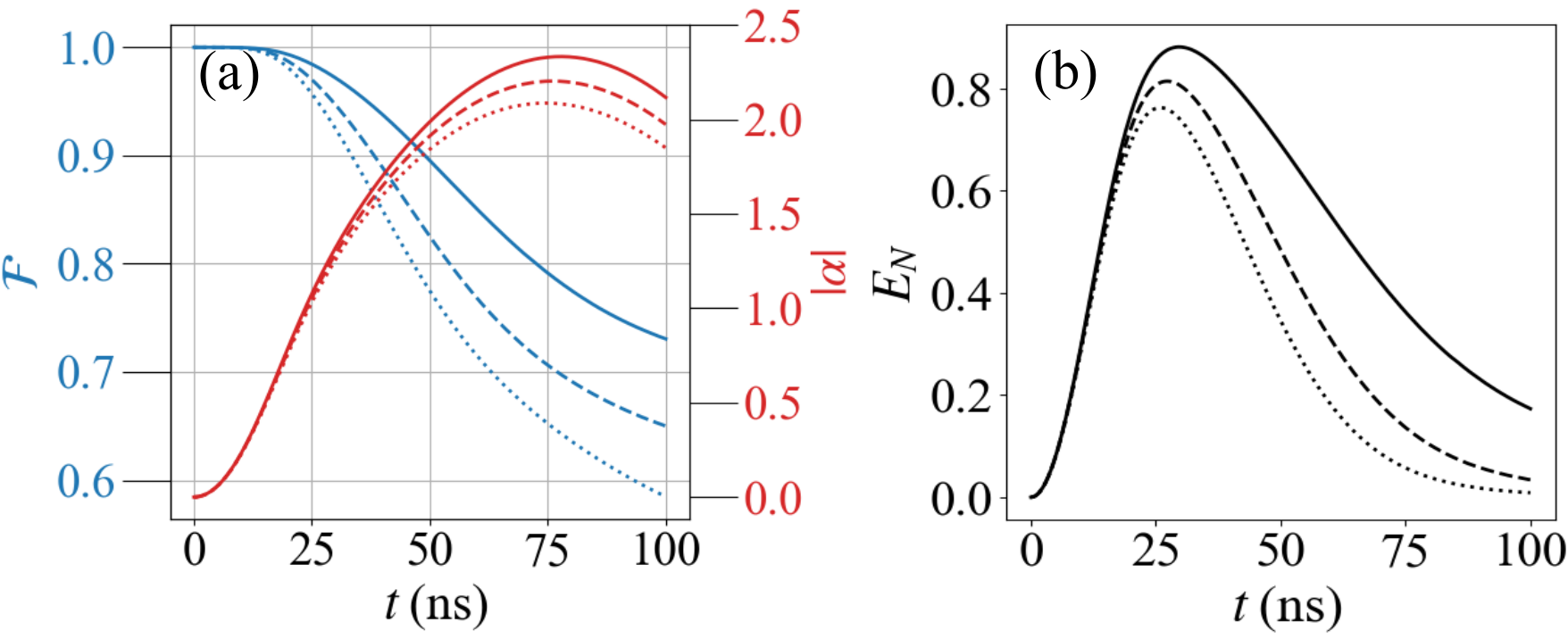}
	\renewcommand{\figurename}{FIG}
	\caption{(a) The fidelity $\mathcal{F}$ (blue curves) and {displacement amplitude $|\alpha(t)|$ (red curves) versus time;} and (b) entanglement (logarithmic negativity $ E_N $) of two magnon modes versus time, for three values of magnon dissipation rate: $ \kappa_{1,2}/2\pi =$ 0.5 MHz (solid), 1 MHz (dashed), and 1.5 MHz (dotted). The other parameters are the same as in Fig.~\ref{Fig4}.} 
	\label{Fig5}
\end{figure}

To further study how the dissipation degrades the two-mode cat state, we plot in Fig.~\ref{Fig5}(a) the fidelity (blue curves) between the generated state and the ideal state ($\ket{ \psi(t) }_{+} $ in Eq.~\eqref{target}) without any dissipation/dephasing,  i.e., $\mathcal{F} = \sqrt{_{+}\bra{ \psi(t) } \rho_{12}  \ket{ \psi(t) }_{+} }$, for different values of the magnon dissipation rate, which is the dominant dissipation channel in our system. It is evident that the fidelity monotonically decreases with time in the short-time limit and it drops more rapidly for a larger dissipation rate. Nevertheless, time is needed to create the cat state, i.e., to realize the magnon displacement. We therefore also plot the displacement amplitude $|\alpha(t)|$ (red curves) in Fig.~\ref{Fig5}(a). Without any dissipation, the expression of $ \alpha(t)$ shows a periodic function of time and gives the same maximum displacement, while the dissipation gradually reduces the achievable maximum displacement over time. This implies an optimal time $t = \pi/g_{m}$ in the first period for creating the ``largest" cat state.
The above results indicate that there is a trade-off between the size of the cat state and the fidelity: the displacement amplitude increases with time (before reaching its maximum), while the fidelity gradually decreases, and the cat state with a larger size is more nonclassical and susceptible to noises introduced through dissipation and dephasing~\cite{PhysRevLett.57.13}.  Taking all these into account, we have chosen experimentally accessible parameters in getting Fig.~\ref{Fig4}.

The two-mode cat is also a non-Gaussian entangled state, which can be revealed by calculating the entanglement between the two original magnon modes.  Figure~\ref{Fig5}(b) shows the logarithmic negativity $E_{N} = \log_2 \, \lVert \rho_{12}^{T_{1}} \rVert_1$ versus time, where $\lVert \rho_{12}^{T_{1}} \rVert_1$ denotes the trace norm of the partial transpose $\rho_{12}^{T_1}$ of the bipartite state $\rho_{12}$~\cite{PhysRevA.65.032314}. Clearly, there is an optimal time to achieve the maximum entanglement due to the aforesaid trade-off effect, and the maximum entanglement reduces with the magnon dissipation rate.   
 
At last, we discuss how to measure the joint two-mode Wigner function $W(\alpha_{1},\alpha_{2})$ in the experiment, which is nontrivial because the two magnon modes are coupled to a single qubit, and both the magnon states are read out via the qubit~\cite{PhysRevLett.130.193603,NatureCommunications2026}.  Nevertheless, this can be realized as follows exploiting the magnon-qubit dispersive coupling.   The dispersive coupling can be achieved by tuning the qubit frequency to a far-detuned dispersive regime via the Autler-Townes effect~\cite{PhysRevLett.130.193603}.
Independent displacement operation onto each magnon mode can be implemented by applying a resonant microwave drive field loaded via a loop antenna. A Ramsey measurement of the qubit is subsequently performed to measure the joint parity of the two magnon modes. 
The joint Wigner function can then be achieved via $ W(\alpha_{1},\alpha_{2}) = \frac{4}{\pi^2} P(\alpha_{1},\alpha_{2})$, with the joint displaced parity  $P(\alpha_{1},\alpha_{2}) = \sum_{m_{1}, m_{2}} ( - 1)^{m_{1} + m_{2}} P_{\alpha_{1}, \alpha_{2}}(m_{1},m_{2})$, where $P_{\alpha_{1}, \alpha_{2}}(m_{1},m_{2}) = \langle m_1,m_2 | D_{1}( -\alpha_{1}) D_{2}( -\alpha_{2}) \rho_{12}  D_{1}(\alpha_{1}) D_{2}( \alpha_{2}) | m_1,m_2 \rangle$ can be obtained from the Ramsey measurement~\cite{science.aaf2941, Nature2022, wsqr-j9f4}.

\section{CONCLUSIONS}\label{S5}

We have presented a protocol to prepare two-mode magnonic Schrödinger cat states based on a cavity-magnon-qubit system. The core of the protocol is to engineer an effective magnon-qubit conditional-displacement interaction, which together with a qubit projective measurement, creates a superposition of four coherent states involving two magnon modes.  Using currently available parameters and including practical dissipations and dephasing, a two-mode cat state with strong nonclassicality and non-Gaussian entanglement can be generated.  The work demonstrates the great potential of the cavity-magnon-qubit system for preparing macroscopic quantum states involving multi-mode magnonic systems, which would enable enhanced quantum detection of dark-matter axions~\cite{wbhn-v1sw,PhysRevLett.124.171801}.

\section*{ACKNOWLEDGMENTS}

This study was supported by Zhejiang Provincial Natural Science Foundation of China (Grant No.~LR25A050001), National Natural Science Foundation of China (Grants No.~12474365 and No.~92265202), and National Key Research and Development Program of China (Grants No.~2024YFA1408900 and No.~2022YFA1405200).

\appendix

\section{Effective Hamiltonian after adiabatic elimination of the cavity}\label{appa}

The Hamiltonian ~\eqref{initial} of the whole system can be divided into two parts:
\begin{equation}
\begin{split}
	H_{\text{1, 0}} =\;\; & \omega_c c^{\dagger} c + \frac{\omega_{Q}}{2} \sigma_z + \omega_1  m_{1}^{\dagger} m_{1} + \omega_2 m_{2}^{\dagger} m_{2},  \\
	H_{\text{1, I}} =\;\; &  g_{1c}\left(m_{1}^{\dagger} c + m_{1} c^{\dagger} \right) + g_{2c} \left( m_{2}^{\dagger} c + m_{2} c^{\dagger} \right) . \\
   & + g_{q c} \left(\sigma_{+}c + \sigma_{-}c^{\dagger} \right)
\end{split}
\end{equation}
To derive the effective couplings among the two magnon modes and the qubit, we adiabatically eliminate the cavity mode by performing {the Fr\"ohlich-Nakajima transformation~\cite{PhysRev.79.845, Nakajima01101955}. }Specifically, we introduce a unitary operator $ U(S) = e^{S} $, which satisfies the relation $ H_{\text{1, I}} + [H_{\text{1, 0}}, S] = 0 $. The operator $ S $ takes the form 
\begin{equation}
\begin{split}
	S =&- \frac{g_{1c}}{\Delta_{1c}}  \left( m_{1} c^{\dagger} - m_{1}^{\dagger} c \right) - \frac{g_{2c}}{\Delta_{2c}}  \left( m_{2} c^{\dagger} - m_{2}^{\dagger} c \right)\\
	&- \frac{g_{qc}}{\Delta_{Qc}}  \left( \sigma_{-} c^{\dagger} - \sigma_{+} c \right).
\end{split} 
\end{equation}
Note that the magnon modes and the qubit are far-detuned from the cavity mode, i.e., $  | \Delta_{Qc} | , |\Delta_{1c}|, |\Delta_{2c}| \gg g_{qc}, g_{1c}, g_{2c} $. {The Hamiltonian becomes}
\begin{equation}\label{A_2}
\begin{split}
  H &= U(S)^{\dagger} H_{1} U (S) \approx H_{\text{1, 0}} + \frac{1}{2} \left[H_{\text{1, I}}, S \right]\\
  &=\left( \omega_{1} - \frac{g_{1c}^2}{\Delta_{1c}}  \right) m_{1}^{\dagger} m_{1} +  \left( \omega_{2} - \frac{g_{2c}^2}{\Delta_{2c}}  \right)   m_{2}^{\dagger} m_{2}\\
  &+ \frac{1}{2} \left( \omega_{Q} - \frac{g_{qc}^2}{\Delta_{Qc}}  \right) \sigma_z\\
  & - \left[ \frac{g_{1c}^2}{\Delta_{1c}}\left( m_{1}^{\dagger} m_{1} - \frac{1}{2}  \right) +\frac{g_{2c}^2}{\Delta_{2c}}\left( m_{2}^{\dagger} m_{2} - \frac{1}{2}  \right) +\frac{g_{Qc}^2}{\Delta_{Qc}}  \sigma_{z} \right]c^{\dagger} c  \\ 
  & - \frac{1}{2} g_{1c} g_{qc}\left( \frac{1}{\Delta_{1c}} + \frac{1}{\Delta_{Qc}} \right)  \left(\sigma_{+}m_{1} + \sigma_{-} m_{1}^{\dagger}  \right) \\
  & - \frac{1}{2} g_{2c} g_{qc}\left( \frac{1}{\Delta_{2c}} + \frac{1}{\Delta_{Qc}} \right) \left( \sigma_{+} m_{2} + \sigma_{-} m_{2}^{\dagger}  \right)\\
  & - \frac{1}{2} g_{1c} g_{2c}\left( \frac{1}{\Delta_{1c}} + \frac{1}{\Delta_{2c}} \right) \left(m_{1}^{\dagger} m_{2} + m_{1} m_{2}^{\dagger} \right).\\
\end{split}
\end{equation}
We assume that the cavity mode remains in the vacuum state, i.e., $ \langle c^{\dagger} c \rangle \approx 0 $, which is the case under a low temperature (e.g., a few tens of mK) and the drive strength we used.  {After neglecting the term containing $ c^{\dagger} c $ in Eq.~(\ref{A_2})}, we obtain
\begin{equation}\label{hhh}
\begin{split}
  H =& \;\; \left( \omega_{1} - \frac{g_{1c}^2}{\Delta_{1c}}  \right) m_{1}^{\dagger} m_{1} +  \left( \omega_{2} - \frac{g_{2c}^2}{\Delta_{2c}}  \right)   m_{2}^{\dagger} m_{2}\\
  &+ \frac{1}{2} \left( \omega_{Q} - \frac{g_{qc}^2}{\Delta_{Qc}}  \right) \sigma_z\\
  & - \frac{1}{2} g_{1c} g_{qc}\left( \frac{1}{\Delta_{1c}} + \frac{1}{\Delta_{Qc}} \right)  \left(\sigma_{+}m_{1} + \sigma_{-} m_{1}^{\dagger}  \right) \\
  & - \frac{1}{2} g_{2c} g_{qc}\left( \frac{1}{\Delta_{2c}} + \frac{1}{\Delta_{Qc}} \right) \left( \sigma_{+} m_{2} + \sigma_{-} m_{2}^{\dagger}  \right)\\
  & - \frac{1}{2} g_{1c} g_{2c}\left( \frac{1}{\Delta_{1c}} + \frac{1}{\Delta_{2c}} \right) \left(m_{1}^{\dagger} m_{2} + m_{1} m_{2}^{\dagger} \right).
\end{split}
\end{equation}
The above effective Hamiltonian describes an excitation-exchange interaction between the qubit and each magnon mode, and between the two magnon modes. For convenience, Eq.~\eqref{hhh} is presented in a simpler form as in Eq.~(\ref{hh}).

\section{Derivation of the unitary evolution operator}\label{appb}

Here, we present a detailed derivation of the unitary evolution operator $ \mathcal{U}(t) $ introduced in Eq.~(\ref{Ut}). The effective Hamiltonian Eq.~(\ref{heff}) (excluding the dark mode) in the interaction picture with respect to $g_{m}U^{\dagger} U$ is given by
\begin{equation}\label{HI}
\begin{split}
	H_{\text{I}}(t) = \frac{g}{\sqrt{2}} \left( U e^{-i g_{m} t} + U^{\dagger} e^{i g_{m} t} \right) \sigma_{x},
\end{split}
\end{equation}
and the Schrödinger equation for the evolution operator is  (setting $ \hbar =1$) 
\begin{equation}
\begin{split}
	i \frac{\mathrm{d}}{\mathrm{d}t} \mathcal{U}(t) = H_{\text{I}}(t) \mathcal{U}(t).
\end{split}
\end{equation}
The evolution operator takes the form of
\begin{equation}
\begin{split}
	\mathcal{U}(t) = \mathcal{T} \exp \left\{ \int_{0}^{t} \left[ - i H_{\text{I}}(\tau) \right]  \mathrm{d}\tau \,  \right\},
\end{split}
\end{equation}
where $ \mathcal{T} $ is the time-ordering operator. Using the Magnus expansion~\cite{PhysRevA.93.033853}, the operator can be expressed as 
\begin{equation}
\begin{split}
	\mathcal{U}(t) = \exp \left[ \Lambda(t) \right],\quad  \Lambda(t) = \sum_{n = 1}^{\infty} \Lambda_{n}(t).
\end{split}
\end{equation} 
The first two terms are given by
\begin{equation}
\begin{split}
	\Lambda_{1}(t) &= \int_{0}^{t} \left[ - i H_{\text{I}}(t_{1}) \right]  \mathrm{d}t_{1} \\
	&= \frac{g}{\sqrt{2}g_{m}} \left[ \left( 1 - e^{i g_{m} t}  \right) U^{\dagger}   - \left(1 - e^{-i g_{m} t} \right) U  \right] \sigma_{x} \\
	& \equiv \left[ \alpha(t) U^{\dagger} - \alpha^{*}(t) U \right] \sigma_{x},\\
	\Lambda_{2}(t) &= \frac{1}{2} \int_{0}^{t} \mathrm{d}t_{1} \int_{0}^{t_{1}} \mathrm{d}t_{2} \left[ - i H_{\text{I}}(t_{1}), - i H_{\text{I}}(t_{2}) \right].\\
	&= i \frac{g^2}{2g_{m}} t - i \frac{g^2}{2 g_{m}^2} \sin g_{m} t \\
	&\equiv  i \Phi(t),
\end{split}
\end{equation}
{where $\alpha(t) =  g[1 - \exp(i g_{m} t)]/(\!\sqrt{2} g_{m})$ and $\Phi(t) = g^2 (g_{m} t - \sin g_{m} t)/(2 g_{m}^2)$.}
Since the commutator of the Hamiltonian at different times reduces to a $ c $-number, all higher-order terms vanish, i.e., $ \Lambda_{n > 2}(t) = 0 $. Consequently, the evolution operator $ \mathcal{U}(t) $ takes the closed form
\begin{equation}
\begin{split}
	\mathcal{U}(t) = e^{i \Phi(t)} e^{ \left[  \alpha(t) U^{\dagger} - \alpha^{*}(t) U \right] \sigma_{x} }, 
\end{split}
\end{equation}
where $ \Phi (t)$ represents to a global phase, and the second exponential term corresponds to a magnon displacement operator conditioned on the eigenvalues of $ \sigma_{x} $.

\section{Derivation of the effective master equation}\label{appc}

Here we provide more details on the derivation of the effective master equation~\eqref{meff} associated with the effective Hamiltonian~\eqref{hheff} from the master equation~\eqref{meq2} associated with the Hamiltonian~\eqref{hh}. The above two Hamiltonians satisfy the relation:
\begin{equation}
\begin{split}
  H_{6} = U_{\text{tot}}  H_{2} U^{\dagger}_{\text{tot}}  - i \frac{\partial U_{\text{tot}} }{\partial t} U_{\text{tot}} ^{\dagger},
\end{split}
\end{equation} 
where $  U_{\text{tot}}  = \exp( - i H_{\text{0}} t) \exp[ - i \omega_{d} ( m_{1}^{\dagger} m_{1} + m_{2}^{\dagger} m_{2} + \sigma_z/2 ) t] $ with $H_{\text{0}} = \Omega \sigma_{x} $. By inserting  the identity $ I = U_{\text{tot}} ^{\dagger} U_{\text{tot}}  $ to the master equation ~\eqref{meq2} and utilizing the relation $  \dot{U^{\dagger} }_{\text{tot}}  U_{\text{tot}}  + U_{\text{tot}} ^{\dagger} \dot{U}_{\text{tot}}  = 0 $, we obtain 
\begin{equation}\label{mefff} 
\begin{split}
	\frac{\mathrm{d} \tilde{\rho}}{\mathrm{d} t} =& - i \left[ H_{\text{6}}, \tilde{\rho} \right] + \frac{\gamma_{q}}{2} \mathcal{L}[\tilde{\sigma}_{-}] \tilde{\rho} + \frac{\gamma_{\phi}}{4} \mathcal{L}[\tilde{\sigma}_{z}] \tilde{\rho} \\
	&+ \frac{\kappa_{1}}{2} \mathcal{L}[\tilde{m}_{1}] \tilde{\rho} + \frac{\kappa_{2}}{2} \mathcal{L}[\tilde{m}_{2}] \tilde{\rho},
\end{split}
\end{equation}
where $ \tilde{\rho} = U_{\text{tot}}  \rho U^{\dagger}_{\text{tot}}  $ and $ \tilde{o} = U_{\text{tot}}  o U^{\dagger}_{\text{tot}}  $. Since $ U_{\text{tot}}  $ does not affect the annihilation operators of the magnon modes, we only need to evaluate the transformed qubit operators in Eq.~(\ref{mefff})
\begin{equation}
\begin{split}
	&U_{\text{tot}}  \sigma_{-} U_{\text{tot}} ^{\dagger} = \frac{1}{2} \left[\sigma_{x} - i \left(  \sigma_{y} \cos 2 \Omega t + \sigma_{z} \sin 2 \Omega t  \right)\right],\\
	&U_{\text{tot}}  \sigma_{+} U_{\text{tot}} ^{\dagger} = \frac{1}{2} \left[\sigma_{x} + i \left(  \sigma_{y} \cos 2 \Omega t + \sigma_{z} \sin 2 \Omega t  \right)\right],\\
	&U_{\text{tot}}  \sigma_{z} U_{\text{tot}} ^{\dagger} = \sigma_{z} \cos(2 \Omega t) - \sigma_{y} \sin(2 \Omega t).
\end{split}
\end{equation}
Substituting the above into each term of the Lindblad superoperator $ \mathcal{L}[\tilde{o}] \tilde{\rho} = 2 \tilde{o} \tilde{\rho} \tilde{o}^{\dagger} - \tilde{o}^{\dagger} \tilde{o} \tilde{\rho} - \tilde{\rho} \tilde{o}^{\dagger} \tilde{o} $, with $ \tilde{o} = \tilde{\sigma}_{-} $, we obtain
\begin{equation}\label{leff}
\begin{split}
	&2 \tilde{\sigma}_{-} \tilde{\rho} \tilde{\sigma}_{+} = \frac{1}{2} \left[
	\sigma_{x} \tilde{\rho} \sigma_{x} + i\sigma_{x} \tilde{\rho} \sigma_{y}\cos 2\Omega t + i\sigma_{x} \tilde{\rho} \sigma_{z} \sin 2 \Omega t \right. \\
	&- i \sigma_{y} \tilde{\rho} \sigma_{x} \cos 2 \Omega t + \sigma_{y} \tilde{\rho} \sigma_{y} \cos^2 2 \Omega t + \sigma_{y} \tilde{\rho} \sigma_{z} \sin \left( 4 \Omega t \right) / 2 \\
	&- i \sigma_{z} \tilde{\rho} \sigma_{x} \sin 2 \Omega t + \sigma_{z} \tilde{\rho} \sigma_{y} \sin \left( 4 \Omega t \right) / 2 + \sigma_{z} \tilde{\rho} \sigma_{z} \sin^2 2 \Omega t \left.
	\right]\\
	&\tilde{\sigma}_{+} \tilde{\sigma}_{-} \tilde{\rho} = \frac{1}{4} \left[
	\sigma_{x} \sigma_{x} \tilde{\rho} - i \sigma_{x} \sigma_{y} \tilde{\rho} \cos 2\Omega t - i \sigma_{x} \sigma_{z} \tilde{\rho} \sin 2 \Omega t \right. \\
	&+ i \sigma_{y} \sigma_{x} \tilde{\rho} \cos 2 \Omega t + \sigma_{y} \sigma_{y} \tilde{\rho} \cos^2 2 \Omega t + \sigma_{y} \sigma_{z} \tilde{\rho} \sin \left( 4 \Omega t \right) / 2 \\
	&+ i \sigma_{z} \sigma_{x} \tilde{\rho} \sin 2 \Omega t + \sigma_{z} \sigma_{y} \tilde{\rho} \sin \left( 4 \Omega t \right) / 2 + \sigma_{z} \sigma_{z}  \tilde{\rho} \sin^2 2 \Omega t \left.
	\right]\\
	&\tilde{\rho} \tilde{\sigma}_{+} \tilde{\sigma}_{-} = \frac{1}{4} \left[
	\tilde{\rho} \sigma_{x} \sigma_{x} - i \tilde{\rho} \sigma_{x} \sigma_{y} \cos 2\Omega t - i \tilde{\rho} \sigma_{x} \sigma_{z} \sin 2 \Omega t \right. \\
	&+ i \tilde{\rho} \sigma_{y} \sigma_{x} \cos 2 \Omega t + \tilde{\rho} \sigma_{y} \sigma_{y} \cos^2 2 \Omega t + \tilde{\rho} \sigma_{y} \sigma_{z} \sin \left( 4 \Omega t \right) / 2 \\
	&+ i \tilde{\rho} \sigma_{z} \sigma_{x} \sin 2 \Omega t + \tilde{\rho} \sigma_{z} \sigma_{y} \sin \left( 4 \Omega t \right) / 2 + \tilde{\rho} \sigma_{z} \sigma_{z} \sin^2 2 \Omega t \left.
	\right]
\end{split}
\end{equation}
Because the driving strength is assumed to be much larger than the qubit dissipation rate, $ \Omega \gg \gamma_{q} $, fast-oscillating terms average out on the dissipation timescale. Under this condition, we apply the RWA and retain only the resonant terms, and then Eq.~(\ref{leff}) simplifies to
\begin{equation}
\begin{split}
	&2 \tilde{\sigma}_{-} \tilde{\rho} \tilde{\sigma}_{+} = \frac{1}{2} \left[
	\sigma_{x} \tilde{\rho} \sigma_{x} + \frac{1}{2} \sigma_{y} \tilde{\rho} \sigma_{y}  + \frac{1}{2} \sigma_{z} \tilde{\rho} \sigma_{z} \right],\\
	&\tilde{\sigma}_{+} \tilde{\sigma}_{-} \tilde{\rho} = \frac{1}{4} \left[
	\sigma_{x} \sigma_{x} \tilde{\rho}  + \frac{1}{2} \sigma_{y} \sigma_{y} \tilde{\rho} + \frac{1}{2} \sigma_{z} \sigma_{z}  \tilde{\rho} \right],\\
	&\tilde{\rho} \tilde{\sigma}_{+} \tilde{\sigma}_{-} = \frac{1}{4} \left[
	\tilde{\rho} \sigma_{x} \sigma_{x} + \frac{1}{2} \tilde{\rho} \sigma_{y} \sigma_{y} + \frac{1}{2} \tilde{\rho} \sigma_{z} \sigma_{z} \right].
\end{split}
\end{equation}
Consequently, the dissipative contribution associated with the qubit relaxation becomes
\begin{equation}
\begin{split}
	\frac{\gamma_{q}}{2} \mathcal{L}[\tilde{\sigma}_{-}] \tilde{\rho} = \frac{\gamma_{q}}{8} \mathcal{L}[ \sigma_{x} ] \tilde{\rho}  + \frac{\gamma_{q}}{16} \mathcal{L}[ \sigma_{y} ] \tilde{\rho}  + \frac{\gamma_{q}}{16} \mathcal{L}[ \sigma_{z} ] \tilde{\rho}.
\end{split}
\end{equation}
Similarly, for the qubit dephasing channel we obtain
\begin{equation}
\begin{split}
	\frac{\gamma_{\phi}}{4} \mathcal{L}[\tilde{\sigma}_{z}] \tilde{\rho} = \frac{\gamma_{\phi}}{8} \mathcal{L}[ \sigma_{y} ] \tilde{\rho}  + \frac{\gamma_{\phi}}{8} \mathcal{L}[ \sigma_{z} ] \tilde{\rho}.
\end{split}
\end{equation}
Combining the above results yields the effective master equation~(\ref{meff}). 

\bibliography{Primary_manuscript}

\end{document}